\documentclass[aps,preprint]{revtex4}

\usepackage[latin1]{inputenc}
\usepackage{graphicx}
\usepackage{hyperref}

\begin{document}

\title{CHARACTERIZING WIDTH UNIFORMITY BY WAVE PROPAGATION}%

\author{Luciano da F. Costa}%
\author{Giancarlo Mutinari}%
\affiliation{Cybernetic Vision Research Group\\
IFSC - University of São Paulo\\
Caixa Postal 369,São Carlos, SP, Brazil\\
13560-970 (luciano@if.sc.usp.br)}

\author{David Schubert}%
\affiliation{Salk Institute, 10010N\\
Torrey Pines Road, La Jolla, USA\\
CA 92037 (schubert@salk.edu)\\}


\begin{abstract}
This work describes a novel image analysis approach to
characterize the uniformity of objects in agglomerates by using
the propagation of normal wavefronts.  The problem of width
uniformity is discussed and its importance for the
characterization of composite structures normally found in physics
and biology highlighted.  The methodology involves identifying
each cluster (i.e. connected component) of interest, which can
correspond to objects or voids, and estimating the respective
medial axes by using a recently proposed wavefront propagation
approach, which is briefly reviewed.  The distance values along
such axes are identified and their mean and standard deviation
values obtained.  As illustrated with respect to synthetic and
real objects (in vitro cultures of neuronal cells), the combined
use of these two features provide a powerful description of the
uniformity of the separation between the objects, presenting
potential for several applications in material sciences and
biology.
\end{abstract}

\maketitle

\section{Introduction}
Structures defined by the agglomeration of basic elements are
often characterized by the presence of objects and voids defined
between the latter.  Figure 1 illustrates such a situation with
respect to neuronal cells grown in vitro on different substrata,
namely fibronection (a) and tissue culture plastic (b). Several
important properties of such structures can be inferred from the
geometry of the objects and voids present in such images, which
has motivated the growing use of imaging methods in physics and in
biology (e.g. ~\cite{Mandelbrot:1983, Hovi:1996, Plotnick:1996,
Einstein:1998}). For example, the migration of nerve and glial
cells and axon elongation in the developing nervous system lead to
different patterns of cell agglomeration and separation
~\cite{Dickson:2002}. In material sciences, the geometrical
properties of the voids have strong influence over the respective
mechanical and electrical properties of the composite and can tell
much about the composite formation process.   The present work
describes a new concept and methodology for characterizing the
width uniformity of the objects or voids in such agglomerates.

\begin{figure}[!h]
  \begin{minipage}[b]{0.45\linewidth}
    \includegraphics[height=4.5cm,width=6cm]{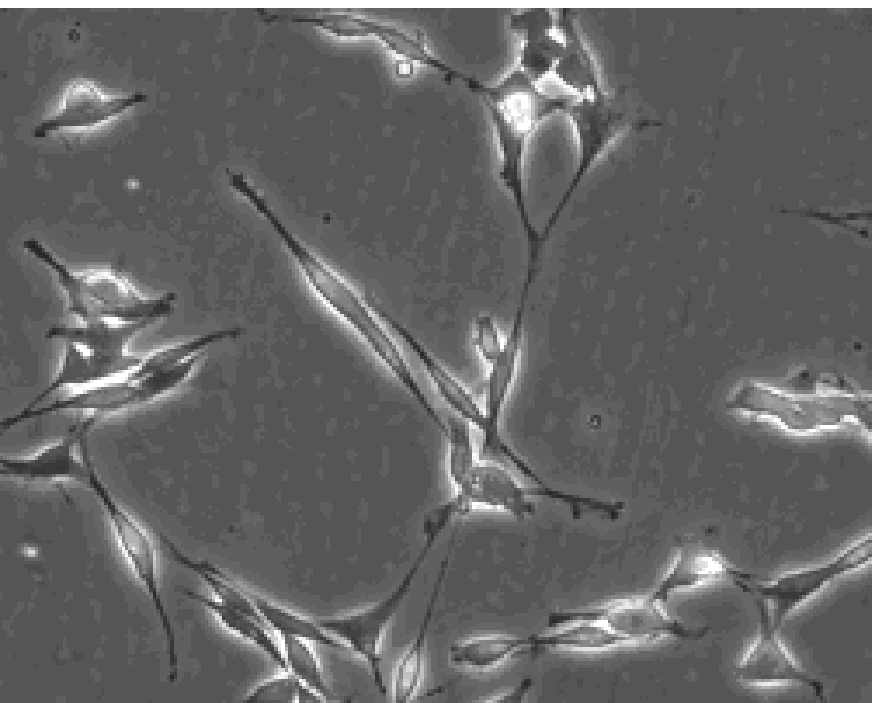}(a)
  \end{minipage} \hfill
  \begin{minipage}[b]{0.45\linewidth}
    \includegraphics[height=4.5cm,width=6cm]{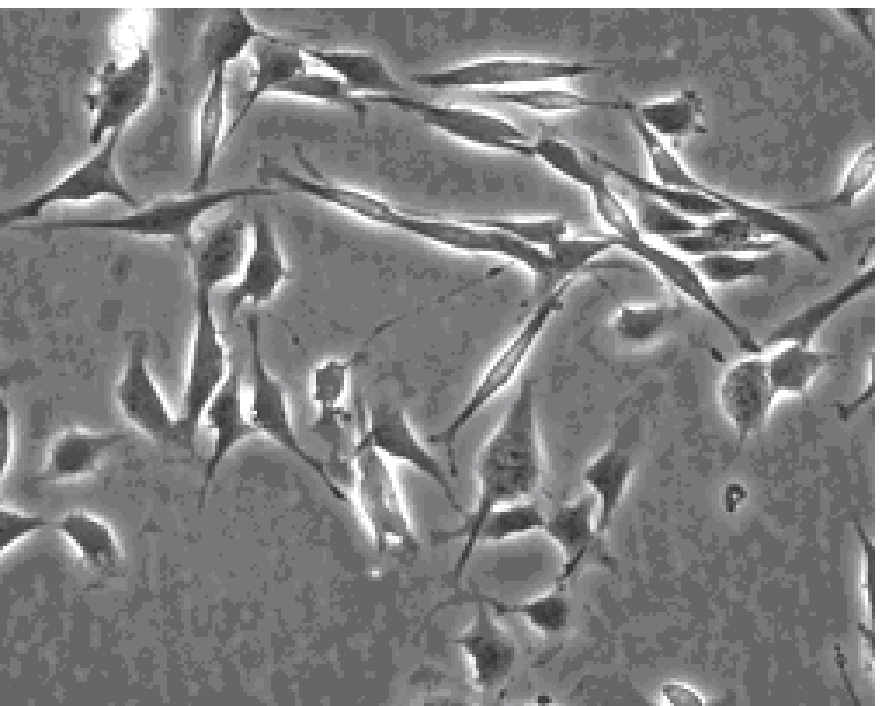}(b)
  \end{minipage}
  \caption{Neuronal cells grown on different substrata, namely fibronectin (a) and tissue culture plastic.}
  \label{fig:cells}
\end{figure}

The first important point is to properly and accurately
characterize what is meant by uniformity of objects or voids.  A
possible interpretation would regard how these elements distribute
themselves along the composite space. We could be interested, for
example, in quantifying their translational invariance, a problem
that could be suitably tackled by using the lacunarity measure
described in ~\cite{Mandelbrot:1983, Einstein:1998, Hovi:1996}.
The alternative problem considered in the present work is to
quantify the uniformity of the width of objects or voids.  Figure
2 illustrates such a situation with respect to two hypothetical
structures chosen for didactical purposes.  The agglomerate in (a)
presents a single void (the black connected region) and objects
(in white) that have similar widths, while the case in (b)
provides a non uniform collection of objects.  The importance of
quantifying the kind of width uniformity seen in (a) is
particularly relevant because it may provide valuable information
about the processes intrinsically related to the composite
formation as well as about the interactions between the basic
elements of the composite. Ideally, it would be interesting to
have a single uniformity index that could be assigned to each
agglomerate image.  The current work proposes an effective
solution to such a problem that can be extended also for analysis
of connected components (i.e. objects or voids) involving
connected holes such as the voids in the two situations in Figure
2.

\begin{figure}[h]
  \begin{minipage}[b]{0.45\linewidth}
    \includegraphics[height=4cm,width=4cm]{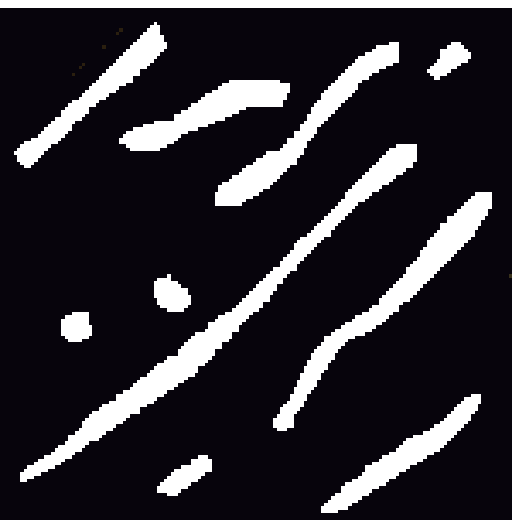}(a)
  \end{minipage} \hfill
  \begin{minipage}[b]{0.45\linewidth}
    \includegraphics[height=4cm,width=4cm]{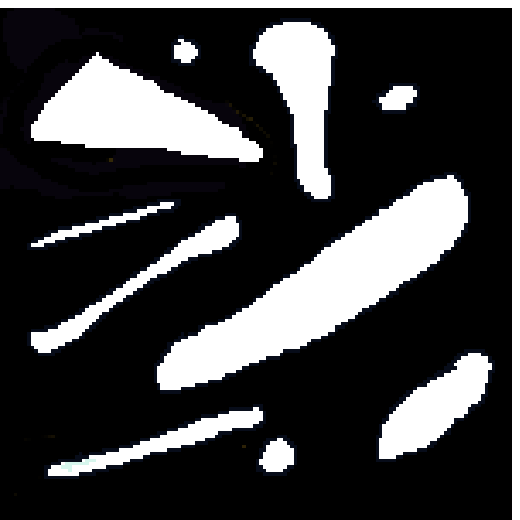}(b)
  \end{minipage}
  \caption{Two types of agglomerates exhibiting different width uniformity of objects (represented in
  white).}\label{fig:agglomerates}
\end{figure}

Firmly based on concepts from mathematics and physics, wavefront
propagation approaches have been extensively used for several
purposes in image analysis and applications (e.g.
~\cite{Osher:1988, Sethian:1999}).  Typically, such methods
involve the numeric propagation, along the object contour normal,
of a wavefront, starting from the object of interest, in such a
way that shocks between portions or individual fronts define the
medial axes of the objects under analysis.  A shock is henceforth
understood as the collision between two different points along the
propagating wavefront(s).  Medial axes are often defined as the
set of points corresponding to the centers of circles maximally
inscribed into the object of interest, which coincide with the
shock points. Also known as skeletons, such axes can be informally
said to correspond to the "middle" part of the objects (see Figure
5). Although alternative approaches to medial axes estimation
(e.g. ~\cite{Ogniewicz:1995}) can be used, the current work
considers the numeric approach reported in (~\cite{Santorini:1999,
Costa:2000, Costa:2001}), which also allows a spatial scale
parameter controlling the degree of detail of the obtained axes.
The basic principle underlying the method for quantifying width
uniformity described in the current paper is the fact that the
distance values along the medial axes provide a natural
characterization of the width distribution of objects even when
they are not straight or present ramifications and holes.  The
potential of the use of the distances along the medial axes is
illustrated in Figure 3. Any object in this figure is
characterized by the peculiar situation that the same distance
value is found along the respective medial axes.   The concept of
width can therefore be extended from straight objects to a more
general set of shapes, including those that are curved or present
branches.  In brief, all shapes in Figure 3(a) can be said to have
the same width. This class of objects has been studied in
~\cite{Attali:1994}] under the name of polyballs.  While such
objects with constant width represent a somewhat limited class of
possible objects, the concept of width can be immediately extended
to more general shapes by considering the distance value along the
object medial axes. Figure 3(b) illustrates such an approach with
respect to a more generic object.  The distances d along the
respective medial axis represent by the dashed line, provide a
natural means to characterize the widths of distribution the
object.

\begin{figure}[h]
  \begin{minipage}[b]{0.45\linewidth}
    \includegraphics[height=4cm,width=4cm]{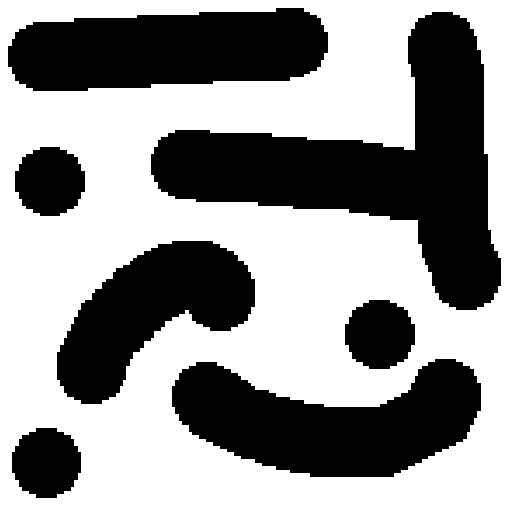}(a)
  \end{minipage} \hfill
  \begin{minipage}[b]{0.45\linewidth}
    \includegraphics[height=3.5cm,width=8cm]{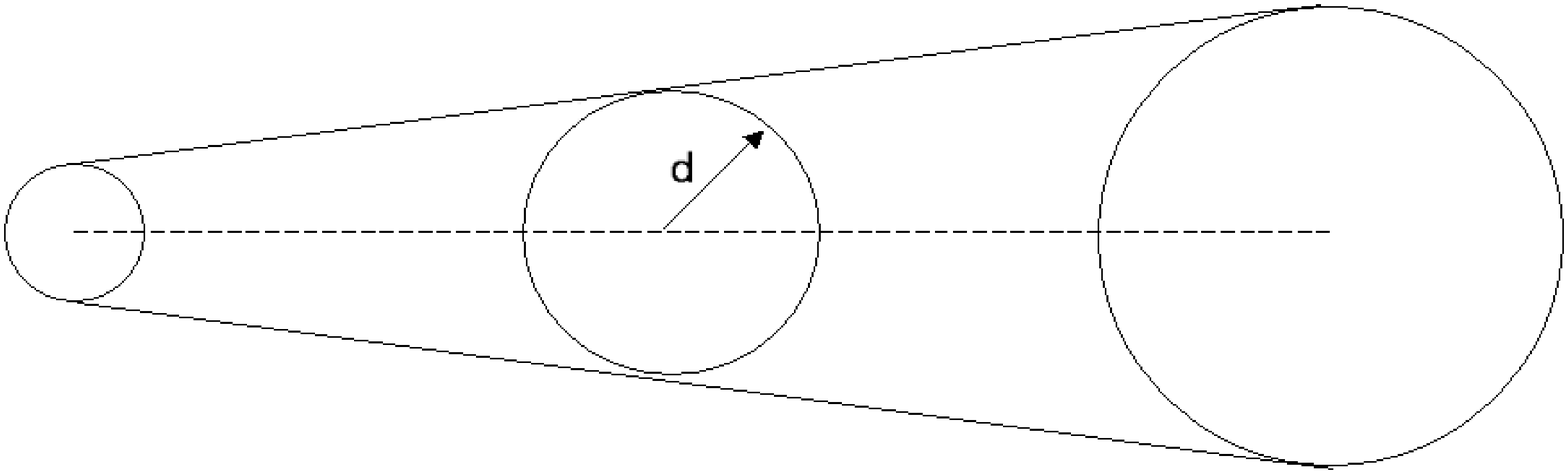}(b)
  \end{minipage}
  \caption{Several objects characterized by the same distances along their respective medial axes(a).
   The generalization of the width of a generic object in terms of the distances along medial
   axes(b).}\label{fig:object}
\end{figure}

Such distributions of distances along the medial axes are
henceforth taken as an accurate indication about the object width.
While the histogram of the distances along the medial axes of all
objects (or voids) in the agglomerate could be taken as a more
comprehensive statistical model for characterizing the object
width, we restrict our attention to its mean and standard
deviation, represented henceforth as $\overline{\emph{w}}$ and
$\sigma_w$, as well as the correlation coefficient $\emph{v}$
between these two measurements. Given its adimensional nature, the
latter provides a particularly interesting candidate for
quantification of the uniformity of the object's width.  As will
be illustrated in this article, these three measures provide
meaningful and comprehensive geometrical characterization of the
objects or voids in agglomerates not only in terms of width
uniformity, but also as far as the size of those objects is
concerned.
This article starts by presenting the methodology
suggested to estimate the medial axes and distances, leading to
the three width features $\overline{\emph{w}}$ , $\sigma_w$, and
$\emph{v}$. Next, the potential of these measures to quantify
width uniformity is illustrated with respect to synthetic and real
data corresponding to agglomerates of neuronal cells grown over
four different types of substrata.

\section{Methodology}
Given an image of an agglomerate characterized by objects over a
background (or any two distinct phases) like that in Figure 2(b),
the first step of the proposed methodology involves identifying
all connected components for the object or voids.  A connected
component is understood as the set of all pixels that are
connected to its 4 or 8 neighbors ~\cite{Costa:2001}. This work
adopts 4-neighborhood. In case we are analyzing voids, a single
connected component is obtained for the agglomerates in Figure 2.
At the same time, a total of 10 connected components are obtained
in case we are interested in the objects.  The extraction of the
connected components can be done by using standard region growing
or inundation algorithms (e.g. ~\cite{Costa:2001}). Let $\emph{P}$
be any of the points of the component to be extracted. Such
methods search for the neighbors $\emph{N}$ of $\emph{P}$ that
have the same pixel value as $\emph{P}$.  The procedure is
repeated for the neighbors $\emph{N}$ until no more neighbors with
the same pixel value are found. Once the connected components have
been obtained, we need to estimate the multiscale medial axes of
each of them.  This procedure is illustrated with respect to one
of the void connected components from the agglomerate in Figure
2(b), shown in Figure 4(a).  First, the border of the connected
component is extracted in clockwise or counterclockwise fashion.
Traditional algorithms for border extraction (e.g.
~\cite{Costa:2001}) are used in the present work.  The output of
such algorithms is a spatially quantified parametric curve defined
by the list of ordered coordinates $\emph{x(i),y(i)}$ , where
$\emph{i=1,2,...,N}$  is the curve parameter corresponding to the
order the border elements are visited by the border extraction
procedure. Figure 4(c) illustrates a portion of the extracted
border of the connected component in (a), parametrized by integer
values corresponding to the sequence in which the border was
followed. The arc length $\emph{s(i)}$ along such a curve is then
accurately estimated by using the spectral method described in
~\cite{Costa:2001}. This method involves obtaining the Fourier
series for the parametric curve defined by the extracted border
and calculating the derivatives by using the property
$\dot{x}\leftrightarrow\Im^{-1}\{\Im(s)G_{\sigma}\}$, where $\Im$
is the Fourier transform and $G_{\sigma}$ is the Fourier transform
of the unit area Gaussian smoothing function
$G_{\sigma}=1/(\sigma\sqrt{2\pi})\exp\{-0.5(\frac{x}{\sigma})^2\}$
with standard deviation $\sigma$. The arc length can be
immediately obtained from the calculated derivatives as
$s(i)=\displaystyle\sum_{k=1}^{i}\sqrt{\dot{x}^2(i)+\dot{y}^2(i)}$
. Additional information about derivative estimation through
spectral methods can be found in ~\cite{Costa:2001}. The value of
$\sigma$ is henceforth fixed at 7 pixels, which minimizes the
effect of the spatially quantized nature of the curve - recall
that in digital images such curves are represented over the
orthogonal lattice - without excessively affecting (smoothing) the
curve.  The estimated arc length for the object in Figure 4(a) is
given in (d). Having obtained the border of the connected
component reparametrized in terms of its arc length, a wavefront
is propagated from the border, unfolding with constant speed and
normal orientation.  The shocks implied by such a traveling
wavefront as a consequence of curvature variations along the
object geometry define the medial axis of the connected component
under analysis.  The methodology proposed in
~\cite{Santorini:1999, Costa:2000, Costa:2001}, and reviewed
below, is used in this work for such a purpose.  The basic idea of
this method is to propagate circular waves centered at each of the
contour elements, according to the Huyghens principle of wave
propagation. Constant radial speed is assumed.  Given that the
objects in digital images are represented over the orthogonal
lattice, which implies specific anisotropies, it is necessary to
consider only those distances that are representable in such
spaces.  For instance, the first such distances are
$0,1,\sqrt{2},2,...$ . Once such distances, which have been called
$\emph{exact distances}$, are pre-calculated (see
~\cite{Costa:2001}) and stored jointly with the respective
relative position vectors with respect to the origin, the
algorithm consists in visiting each of the border elements for
each subsequent exact distance, and updating the empty cells
indexed by the respective relative positions with the current
distance value or the respectively associated arc length. In the
former case, the resulting image corresponds to the distance
transform of the object, which associates to each surrounding
pixel the smallest Euclidean distance to the object. In case the
arc length is updated, we get a labeled image that can be further
processed in order to obtain label differences (see
~\cite{Costa:2001, Costa:2000}) corresponding to the intensity of
the shocks between different portions of the traveling wavefront.
Figures 4(e) and (f) illustrate the distance transform and
propagated labels obtained for the object in (a).

\begin{figure}[ht]
  \begin{minipage}[b]{0.45\linewidth}
    \includegraphics[height=3cm,width=5cm]{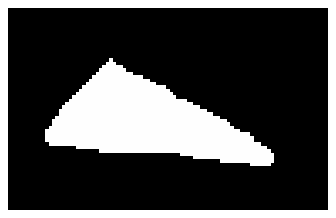}(a)
  \end{minipage} \hfill
  \begin{minipage}[b]{0.45\linewidth}
    \includegraphics[height=3cm,width=5cm]{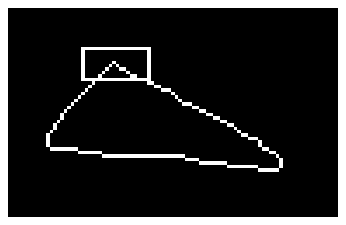}(b)
  \end{minipage}\\
  \begin{minipage}[b]{0.45\linewidth}
    \includegraphics[height=3.5cm,width=4cm]{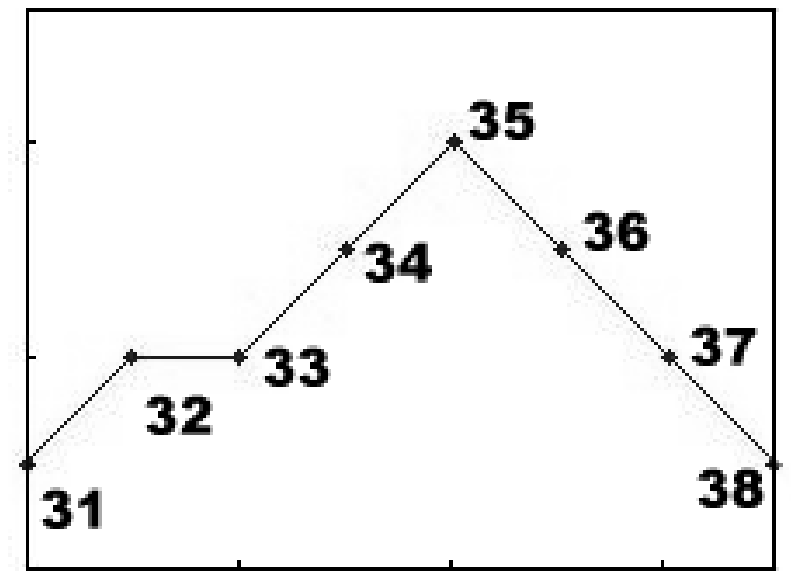}(c)
  \end{minipage} \hfill
  \begin{minipage}[b]{0.45\linewidth}
    \includegraphics[height=4cm,width=5cm]{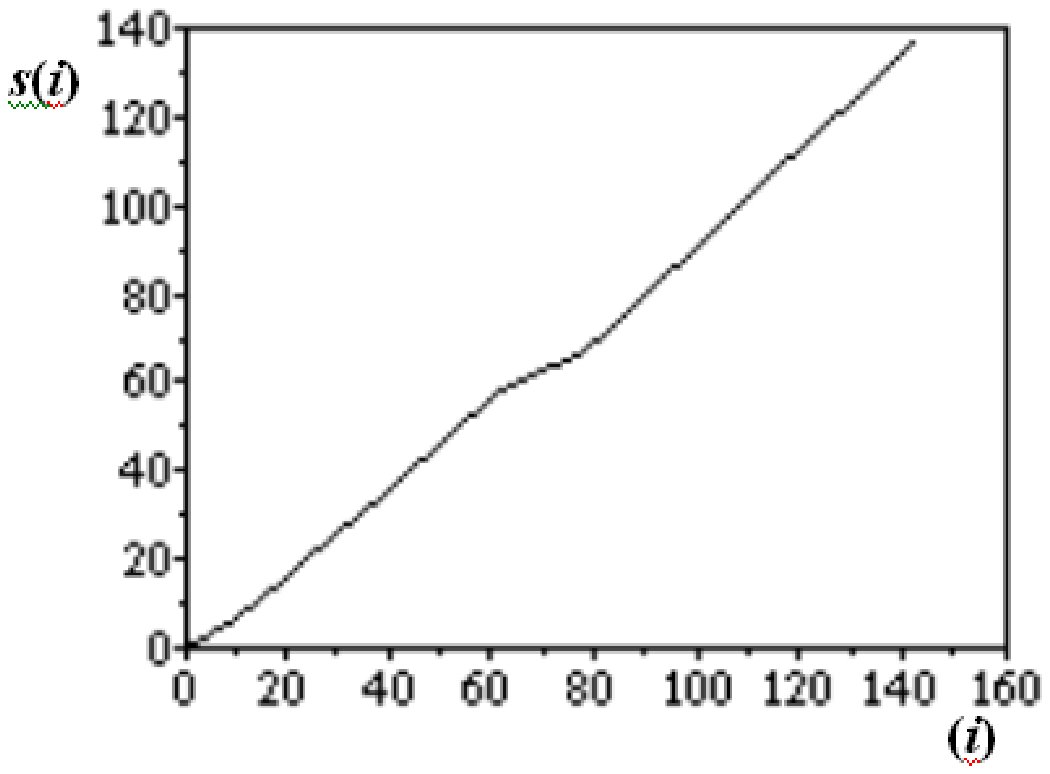}(d)
  \end{minipage}\\
  \begin{minipage}[b]{0.45\linewidth}
    \includegraphics[height=5cm,width=6cm]{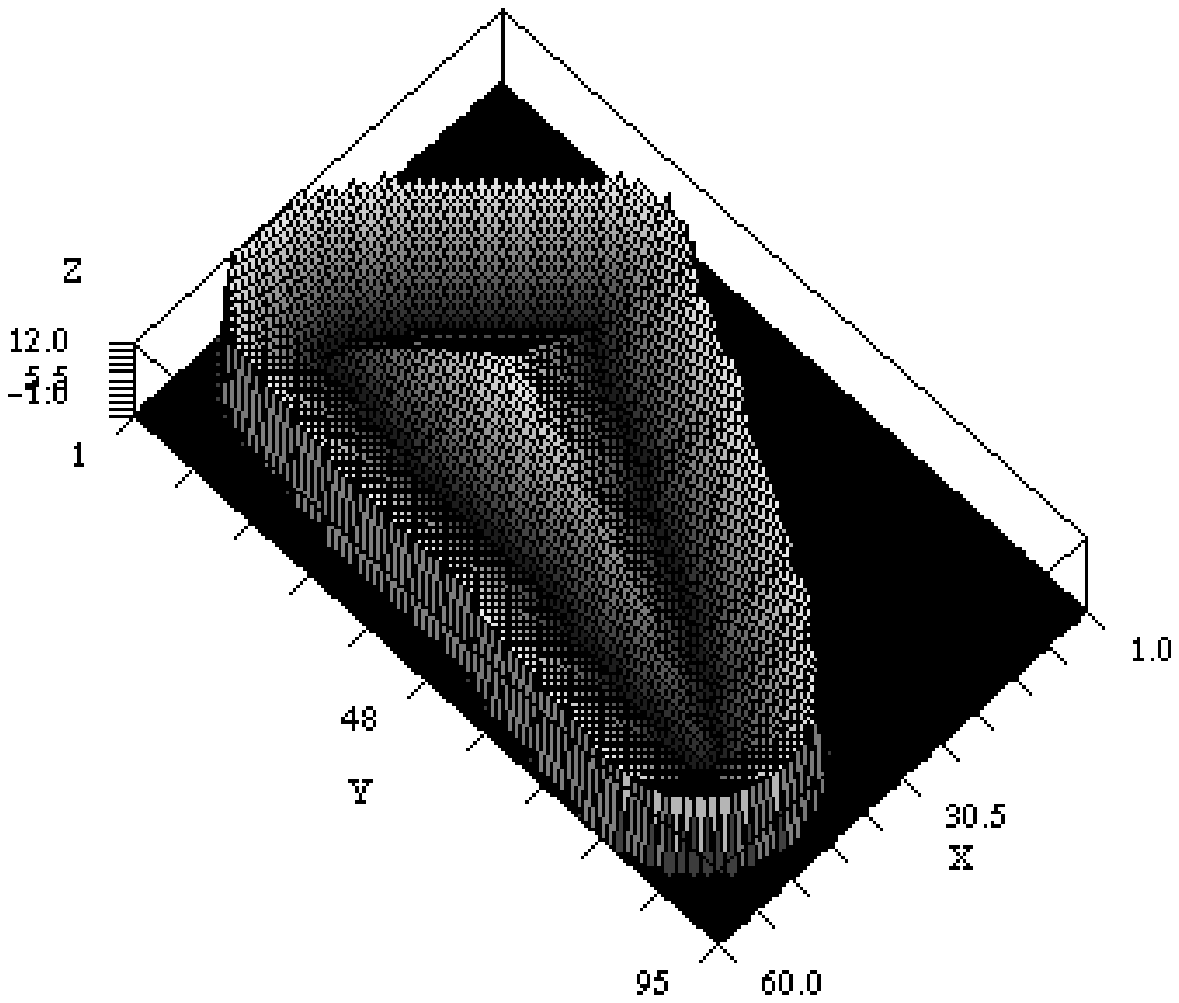}(e)
  \end{minipage} \hfill
  \begin{minipage}[b]{0.45\linewidth}
    \includegraphics[height=4cm,width=6cm]{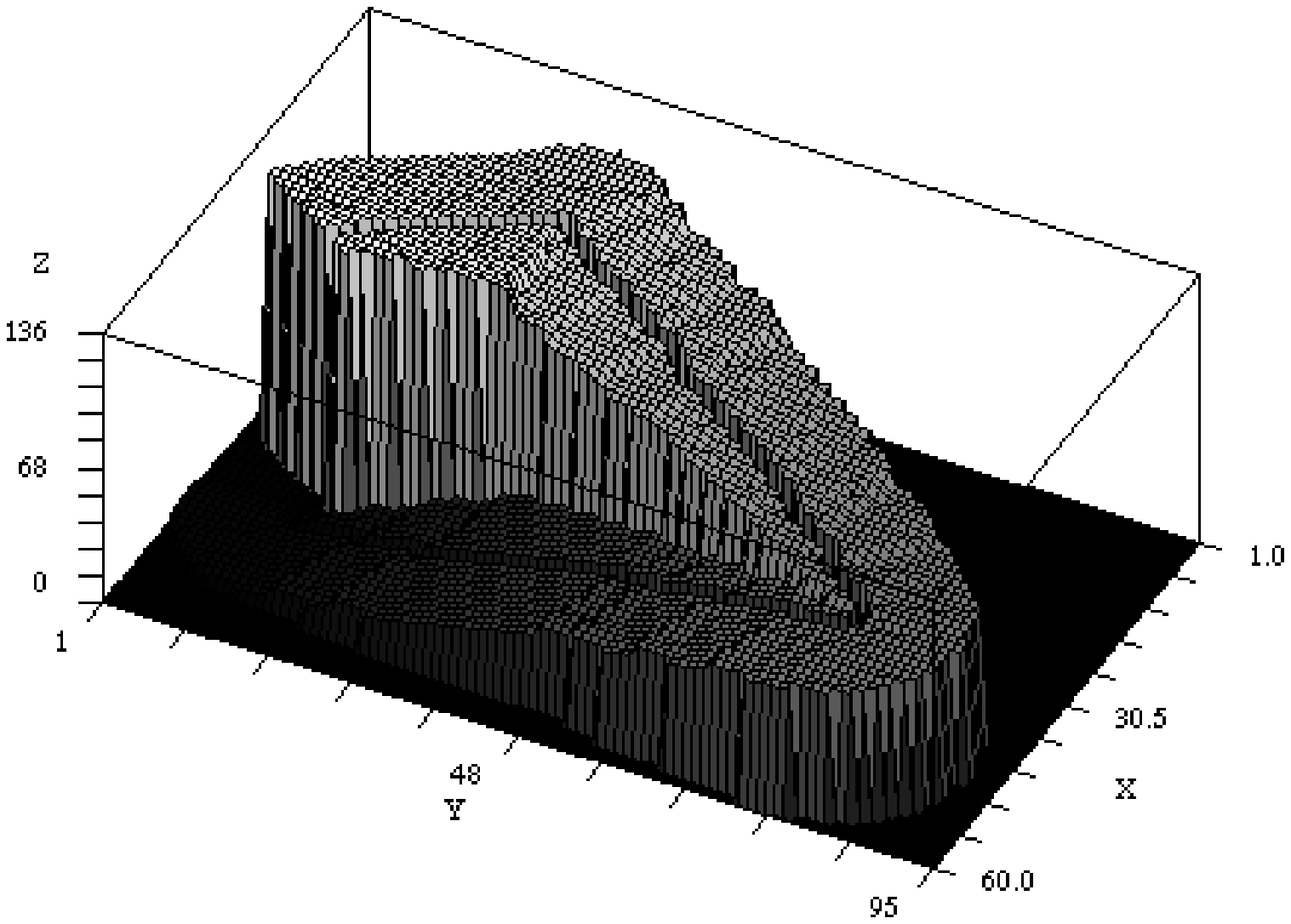}(f)
  \end{minipage}
  \caption{Illustration of the multiscale medial axes estimation by using level
  sets.  The original connected component(a), its extracted border(b), parameter
  values along a piece of such a border(c), arc-length along the parametrized contour(d),
   distance transform(e), and propagated
   labels(f).}\label{fig:labels}
\end{figure}

By thresholding the difference image with different values
$\emph{T}$, it is possible to obtain a family of medial axes with
varying degrees of detail. The higher the value of $\emph{T}$, the
more the details are filtered out.  The application of the above
methodology for medial axes estimation to the problem of width
uniformity involves selecting a suitable value of $\emph{T}$ so as
to remove unimportant small scale detail such as vertices along
the object.  Figure 5 illustrates several medial axes obtained for
varying values of $\emph{T}$. The specific choice of the threshold
value should take into account the resolution and specific demands
imposed by each application.

\begin{figure}[h]
  \begin{minipage}[b]{0.45\linewidth}
    \includegraphics[height=3.5cm,width=5cm]{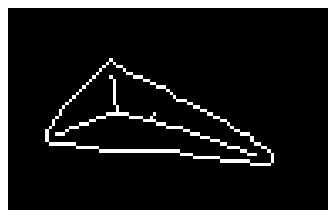} (a)
  \end{minipage} \hfill
  \begin{minipage}[b]{0.45\linewidth}
    \includegraphics[height=3.5cm,width=5cm]{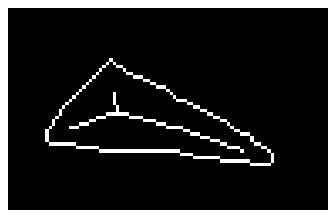} (b)
  \end{minipage}
\end{figure}

\begin{figure}[ht]
  \begin{minipage}[b]{0.45\linewidth}
    \includegraphics[height=3.5cm,width=5cm]{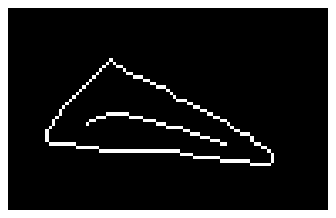} (c)
  \end{minipage} \hfill
  \begin{minipage}[b]{0.45\linewidth}
    \includegraphics[height=3.5cm,width=5cm]{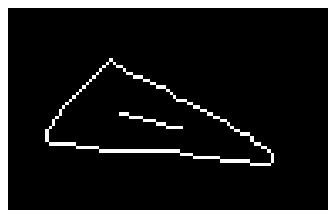} (d)
  \end{minipage}
  \caption{Medial axes obtained for T = 5 (a), 10 (b), 20 (c) and 50
  (d).}\label{fig:medial}
\end{figure}

If connected component presents holes, as is the case with the
composites in Figure 1, each hole has to be identified by using
border extraction algorithms and labeled individually with
specific labels (i.e. a single distinct label is assigned to each
border).  By applying the wave propagation method to such borders,
the generalized Voronoi diagram of the borders is obtained, and
the multiscale medial axes are then calculated inside each of the
Voronoi cells by using the same method described above
~\cite{Falcao:2002}. The generalized Voronoi diagram corresponds
to a partition of the image space in such a way that all points in
each obtained region are closest to the respective object (one of
the image connected component) than to any other object.  Recall
that the distance between a point and an object is defined as the
smallest distance between the point and any of the points
composing the object. Figure 6 illustrates the generalized Voronoi
diagram and the distance values along the borders (fixed-scale
medial axes) determined by the Voronoi cells.

\begin{figure}[h]
  \begin{minipage}[b]{0.45\linewidth}
    \includegraphics[height=5cm,width=5cm]{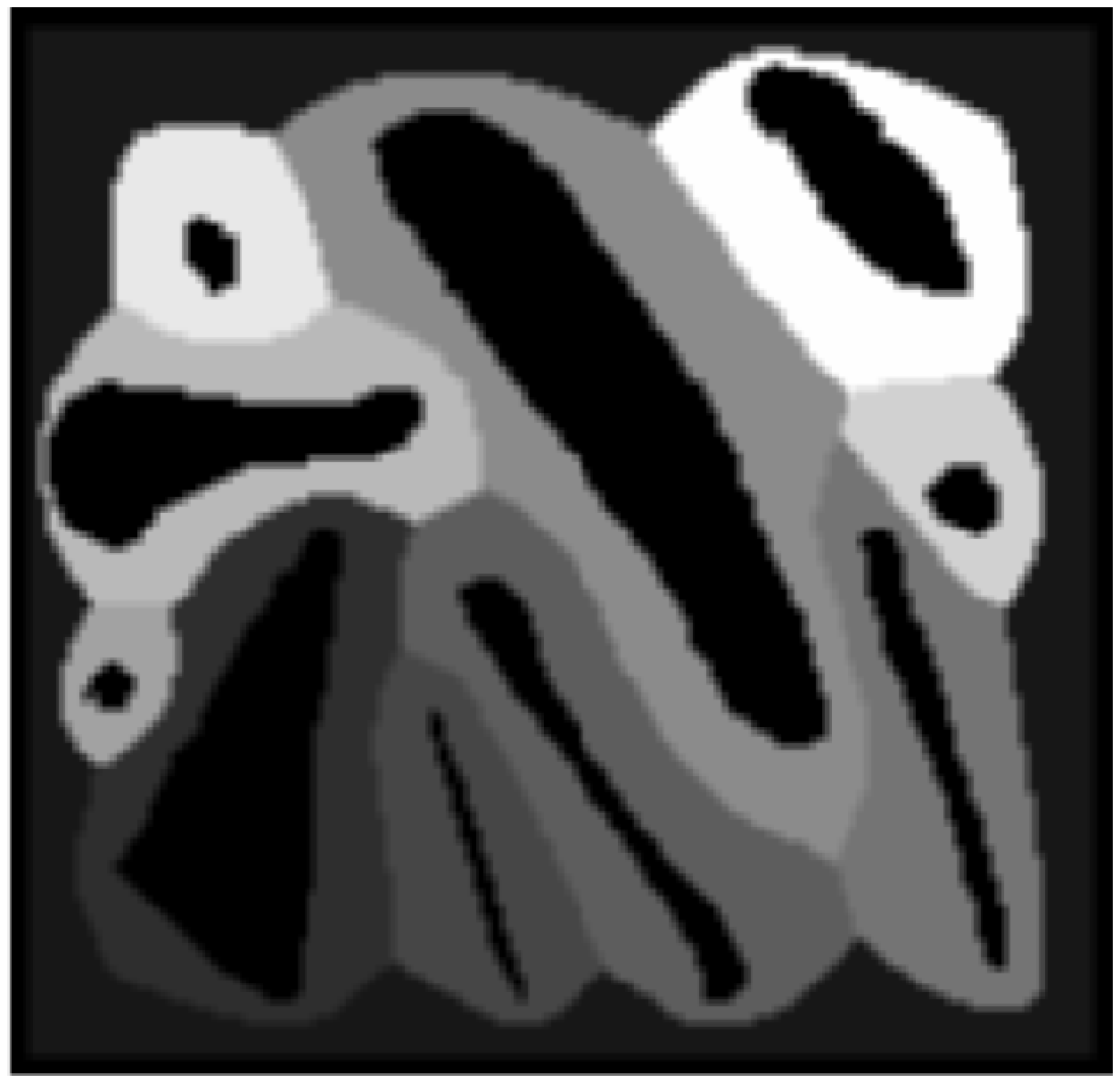} (a)
  \end{minipage} \hfill
  \begin{minipage}[b]{0.45\linewidth}
    \includegraphics[height=5cm,width=5cm]{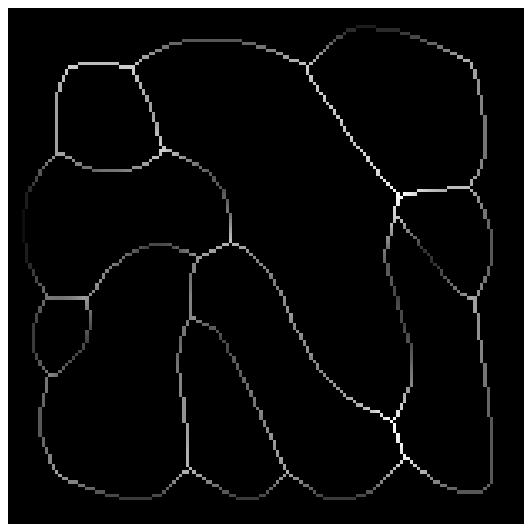} (b)
  \end{minipage}
  \caption{The generalized Voronoi diagram (b) for the composite in Figure 1(b)
   and the respective distance transform values (a) obtained along the Voronoi
   borders.}\label{fig:voronoi}
\end{figure}

At this point, it is possible to derive a distance weighted medial
axis by assigning to each point of the selected medial axes the
respective distance transform values.  The set of distance values
found along the medial axis are henceforth called \emph{medial
axis distances} or \emph{skeleton distances}.  The methodology for
width uniformity characterization involves the consideration of
the distance values along all the medial axes in the agglomerate.
The mean and standard deviation of the distances over every medial
axis point of the object are then estimated.  Table I gives these
values, as well as the correlation coefficient, for the two images
in Figure 1 considering objects and voids.

\begin{table}[ht]
  \centering
  \large
  \caption{The mean, standard deviation and correlation coefficient obtained for the voids
   and objects with respect to the two composites in Figure 1\\}
  \begin{tabular}{|c||c|c|c|}
    \hline & $\overline{\emph{w}}$ & $\sigma_w$ & $\emph{v}$\\\hline
    \hline Fig. 1(a), objects & 2.06 & 0.62 & 0.30\\
    \hline Fig. 1(a), voids & 6.57 & 3.71 & 0.57\\
    \hline Fig. 1(b), objects & 4.37 & 2.29 & 0.52\\
    \hline Fig. 1(b),voids & 6.09 & 3.25 & 0.54\\
    \hline
  \end{tabular}
\end{table}

Several interesting facts result, or can be inferred
from such measures.  To start with, the mean width
$\overline{\emph{w}}$ obtained for the objects (cells) in Figure
2(a) is less than half of the value obtained for the objects in
Figure 2(b), which is in full agreement with the visual
interpretation of those images.  At the same time, the mean widths
obtained for the voids are similar in both images.  The standard
deviations obtained for the objects in those images are also in
full agreement with the fact that the width dispersion of the
objects in (a) is much smaller than that characterizing (b).
Similar values are obtained for the voids. The coefficient of
variation $\emph{v}$ provides an adimensional characterization of
the observed width dispersion differences for the composites in
Figure 2.

\section{Application to Neuronal Agglomerates}
During the development of the nervous system the distribution and
connections of cells is determined by a variety of factors,
including the adhesion of cells and axons to extracellular matrix
material such as laminin and fibronectin ~\cite{Dickson:2002}
little is know, however, about the determinants of cell shape.
Some aspects of cell-matrix adhesions and their relationship to
cell shape can be studied in vitro by plating nerve cells derived
from a single common precursor cell (clonal cells) on different
surfaces and asking how the different surfaces regulate the
morphology of the cell.  Figure 7 shows that there are great
differences in the shape of a clonal nerve cell line called B103
grown on polylysine, laminin, fibronection and tissue culture
plastic.  400 X 319 pixel images of the cultured cells were
obtained by using a Leitz DMIRB inverted microscope and an Openlab
(Leica) image acquisition system connected to a personal computer.
Cell shape can be assigned a numerical value and questions asked
about how the shape of cells grown on the different substrata
differ from each other. The original gray-level images, such as
those illustrated in Figure 1, were segmented ~\cite{Costa:2001}
in order to separate the cells from the substrate, yielding binary
images where the voids are marked as "0" and the cells (or
objects) as "1".  Attention is concentrated on the analysis of
void uniformity.  The segmentation was performed by automated
segmentation (thresholding) followed by manual editing of the
obtained structures in order to remove artifacts such as the
presence of spurious particles and shadows. Figure 7 presents
examples of binary images obtained by the segmentation methodology
described above.  Each of the connected voids were isolated by
using a standard region growing approach ~\cite{Costa:2001}, and
the multiscale medial axes were obtained by using the level
set-based approach as described in Section 2.  Specific medial
axes were obtained by selecting the spatial scale to 10.

\begin{figure}[ht]
  \begin{minipage}[b]{0.45\linewidth}
    \includegraphics[height=5cm,width=5cm]{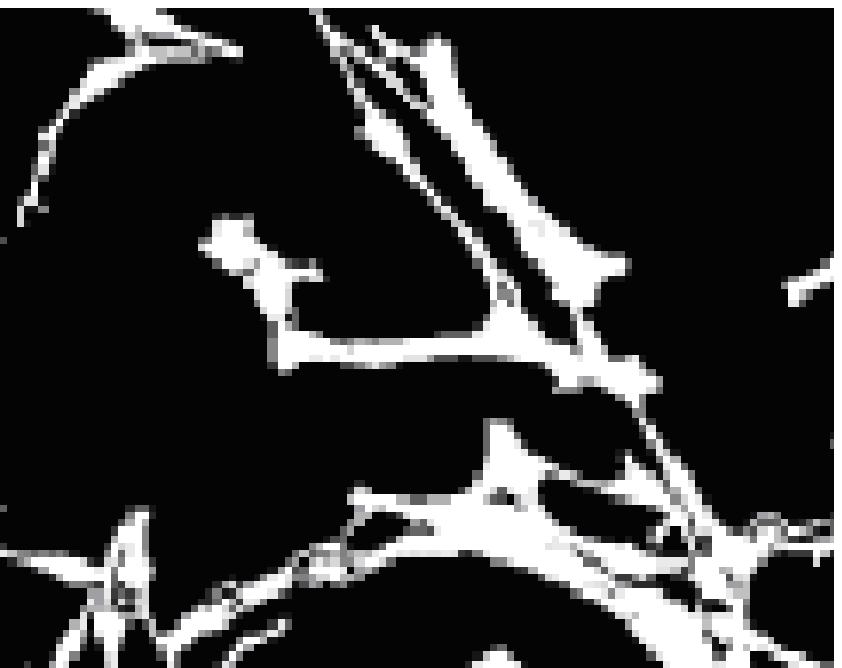} (a)\\
    \centering 20: $\overline{\emph{w}}=7.94$ and
    $\sigma_w=3.62$
  \end{minipage} \hfill
  \begin{minipage}[b]{0.45\linewidth}
    \includegraphics[height=5cm,width=5cm]{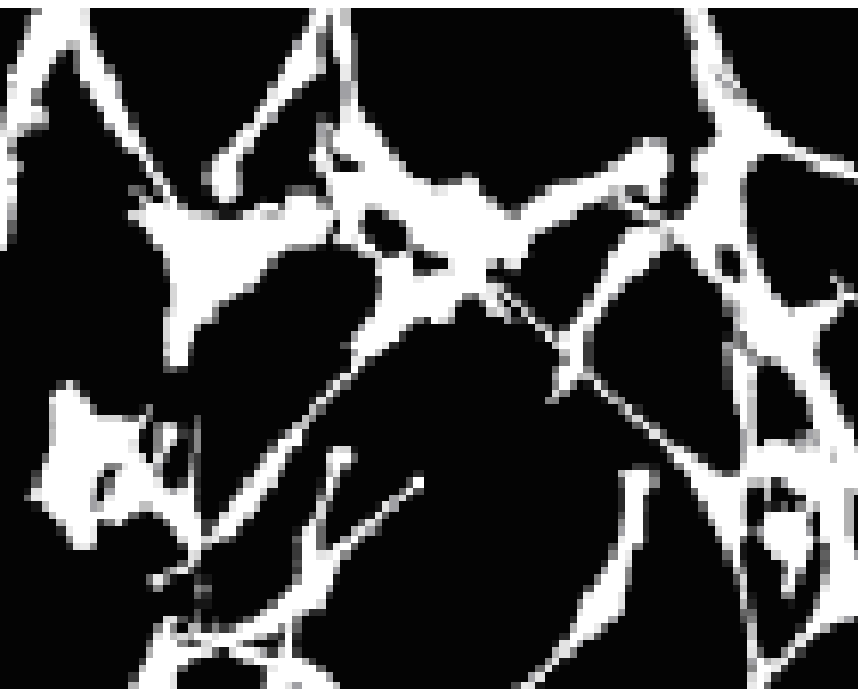} (b)\\
    \centering 32: $\overline{\emph{w}}=5.51$ and
    $\sigma_w=2.67$
  \end{minipage}
\end{figure}
\clearpage
\begin{figure}[ht]
  \begin{minipage}[b]{0.45\linewidth}
    \includegraphics[height=5cm,width=5cm]{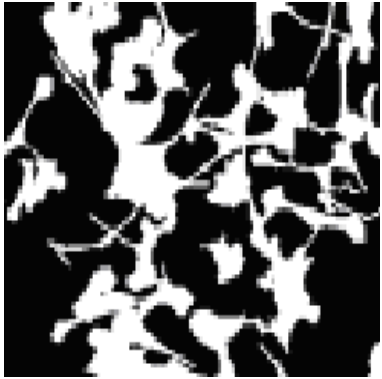} (c)\\
    \centering 56: $\overline{\emph{w}}=13.35$ and
    $\sigma_w=5.26$
  \end{minipage} \hfill
  \begin{minipage}[b]{0.45\linewidth}
    \includegraphics[height=5cm,width=5cm]{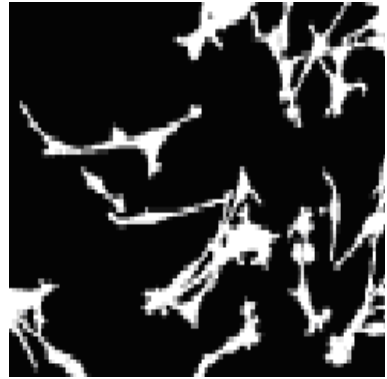} (d)\\
    \centering 70: $\overline{\emph{w}}=10.09$ and
    $\sigma_w=5.53$
  \end{minipage}
  \caption{Binary images (after segmentation of the neuronal cells,
  shown in white, the voids are consequently black) showing examples
  of neuronal agglomerates grown over different substrata and respective
  values of $\overline{\emph{w}}$ and $\sigma_w$.(a): fibronectin; (b): laminin;
  (c): tissue culture plastic;  (d): polylysine.}\label{fig:binary}
\end{figure}

Figure 8 shows the phase space (or scatterplot) obtained by
considering the mean and standard deviation of the void medial
axes distances.  The first important feature in this plot is the
relatively strong overall correlation coefficient, namely 0.86,
between the two considered measures.  While an almost linear
separation was obtained between classes 3 (laminin, triangles) and
4 (polylysine, circles), the other two classes led to more
dispersed, overlapped clusters.  Regarding the mean value of
$\overline{\emph{w}}$, calculated for each class, we have
$mean(\overline{\emph{w}}_3)<mean(\overline{\emph{w}}_2)<mean(\overline{\emph{w}}_1)$
, indicating that the cells grown over these types of substrate
tended to be progressively less packed. This is probably a
reflection of a competition between cell-cell and cell-substratum
adhesion on each surface cells that adhere better to themselves
than to the surface on which they are grown tend to aggregate or,
in the case of neurons, form fascicles ~\cite{Schubert:1991}.
Therefore, cells grown on more adhesive surfaces such as tissue
culture plastic would tend to be spread out on the surface, while
those on less adhesive substrate would form clumps as observed in
Figure 7.
The correlation coefficient obtained for each of the
four clusters are respectively 0.91, 0.73, 0.84 and 0.87,
indicating substantially distinct uniformities characterizing each
substrate. Whereas the voids in class 1 are characterized by the
highest width uniformity, those in class 2 exhibit the most
heterogeneous widths.

\begin{figure}[ht]
  \includegraphics[height=9cm,width=15cm]{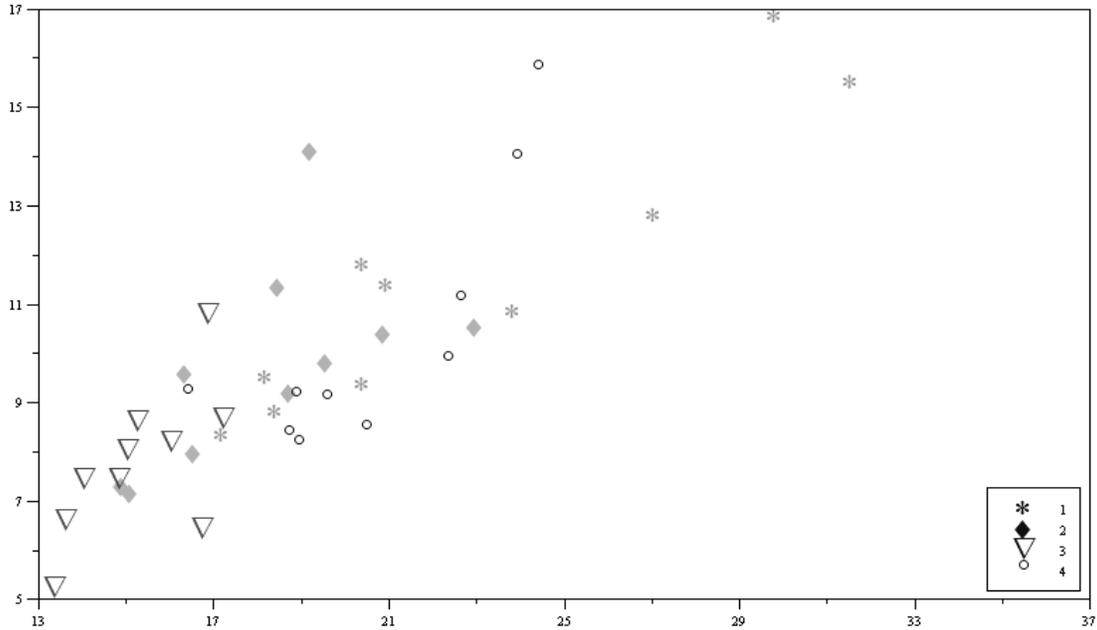}
  \caption{Scatterplot showing the distribution of the neuronal agglomerates.
  The clusters obtained for class 3 and 4 are clearly linearly separable,
  while the other clusters are characterized by substantial
  overlap.}\label{fig:scatter}
\end{figure}

\section{Concluding Remarks}
This article has reported a novel and comprehensive approach to
characterizing width uniformity in objects and voids typically
found in images of agglomerates.  The methodology is based on the
objective characterization of "object width" in terms of the
distances found along its medial axes, which are obtained by using
a wave front propagation scheme. In this way, it becomes possible
to speak objectively about the width of a wide class of shapes,
including those that are curved or present branches.  Such an
approach to width characterization is further enhanced by the fact
that most objects can be approximated or decomposed into varying
radius polyballs.
An accurate numerical approach has been adopted
for the estimation of the medial axis distances that involves
object isolation, border extraction, arc length reparametrization
through a spectral approach, and the calculation of multiscale
medial axes by using wavefront propagation.  Several of these
methods were only recently described in the literature and are
applied in combined fashion for width estimation for the first
time in the present work.  The potential of the proposed framework
was illustrated and corroborated with respect to synthetic
agglomerate images and real data concerning neuronal cells grown
on different substrata.  By examining a clonal nerve cell line
grown on different protein substrata it was possible, using this
new methodology, to quantify the morphological differences caused
by the different substrata. It is shown that when nerve cells are
grown on different surfaces, there is a dramatic correlation
between the shape of the obtained voids and the tendency of the
cells to associate between themselves.  The use of the above
methodology combined with complementary agglomerate measures such
as the fractal dimension and lacunarity provide potential for a
rich and comprehensive characterization of the geometry of the
shapes and agglomerates of both biological and physical objects.\\

\bibliographystyle{plain}
\bibliography{artigo_Gian}

\end{document}